\documentclass[showpacs,amsmath,amssymb,showkeys,onecolumn,a4paper,preprint,nofootinbib,superscriptaddress,floatfix]{revtex4-1}

\usepackage{amsfonts}
\usepackage{amssymb,amscd}
\usepackage{hyperref}
\usepackage{xcolor}
\usepackage{comment}
\usepackage{listings}
\usepackage{wrapfig}
\usepackage{graphicx}
\usepackage{lmodern}
\usepackage[T1]{fontenc}
\usepackage[utf8]{inputenc}	
\usepackage{slashed}

\hypersetup{
    colorlinks=true,                
    breaklinks=true,
    urlcolor= black,                
    linkcolor= blue,                
    bookmarksopen=false,
    filecolor=black,
    citecolor=blue,
    linkbordercolor=blue,
    pdfborder = {0 1 0}
}

\newcommand{\zp}{Z^{\prime}}
\newcommand{\zmp}{Z_{\mu}^{\prime}}
\newcommand{\zpp}{Z^{\prime\mu}}

\newcommand{\Mdf}{M_{Z^{\prime}}}
\newcommand{\mx}{m_{\chi}}
\newcommand{\xs}{\sigma}
\newcommand{\xstot}{\sigma_{\text{tot}}}
\newcommand{\xstotl}{\sigma^{\prime}_{\text{tot}}}
\newcommand{\hxstotl}{\hat{\sigma}^{\prime}_{\text{tot}}}
\newcommand{\hxstot}{\hat{\sigma}_{\text{tot}}}
\newcommand{\dif}{\mathrm{d}}
\newcommand{\ee}{e^{+}e^{-}}
\newcommand{\MET}{\slashed{\vec{E}}_T}
\newcommand{\gx}{g_{\chi}}
\newcommand{\grx}{g_{r\chi}}
\newcommand{\glx}{g_{l\chi}}
\newcommand{\gr}{g_{r}}
\newcommand{\gl}{g_{l}}
\newcommand{\grl}{g_{r/l}}
\newcommand{\xg}{x_{\gamma}}
\newcommand{\xgmin}{x_{\gamma}^{\rm{min}}}
\newcommand{\sigmav}{\langle\sigma v\rangle}

\begin{document}

\title{Investigation of spin-dependent dark matter in mono-photon production at high-energy colliders}


\author{G. Gil da Silveira}
 \email{gustavo.silveira@cern.ch}
\affiliation{CERN, PH Department, 1211 Geneva, Switzerland}
\affiliation{Group of Analysis and Simulation of Particles (GASP), IF-UFRGS \\
Caixa Postal 15051, CEP 91501-970, Porto Alegre, RS, Brazil}

\author{M. S. Mateus Jr.}  \email{msmateusjr@gmail.com}
\affiliation{Group of Analysis and Simulation of Particles (GASP), IF-UFRGS \\
Caixa Postal 15051, CEP 91501-970, Porto Alegre, RS, Brazil}

\begin{abstract}
Many theories about dark matter have emerged due to its strong theoretical appeal in explaining astrophysical phenomena. However, experimental and theoretical particle physics have yet not provided evidence that dark matter is part of the observable Universe. Our work aims to investigate the interaction between Standard Model (SM) fermions and different species of dark matter (DM) particles in high-energy collisions through interaction of a new massive vector mediator, $\zp$. The production of scalar and fermion DM pairs via fermion annihilation into the new vector boson is investigated near a resonance, where a SM signal from hard photon emission is considered as initial state radiation, namely a mono-photon production. Values of coupling constants between the DM and the SM particles are mapped in contrast to the Planck satellite data for thermal relic density DM computed in the correct framework for the relic density near a resonance, where a weaker suppression of the relic density is expected. We show for the CLIC and LHC kinematic regimes that certain mass ranges and coupling constants of these DM particles are in agreement with the expected relic density near a resonance and are not excluded by collider and astrophysical limits.
\end{abstract}

\maketitle

\section{Introduction}\label{intro}

The Standard Model (SM) of the elementary particle interactions has been tested for a variety of phenomena in particle physics at great precision. Nevertheless, there is no (currently) particle in the SM that satisfies the characteristics of the dark matter (DM), i.e., a suitable candidate to explain astrophysical phenomena in the Universe. Neutrinos, for example, known to have non-zero mass \cite{ParticleDataGroup:2022pth}, would be an ideal candidate for DM, however their mass is too small to account for large structure formation \cite{Bertone:2004pz,Bauer:2017qwy}. It is reasonable, therefore, to conceive extensions of the SM that could include new particles and interactions that are consistent with an even more complete description of nature. Several studies have been proposed to investigate the DM and to decipher its origin and nature \cite{Bertone:2004pz,Klasen:2017jiy,Bauer:2017qwy,Duerr:2018mbd,Schumann:2019eaa,PerezdelosHeros:2020qyt,Catena:2013pka,Arcadi:2019lka}, where distinct approaches aim to understand how DM interacts, with itself and with the SM particles, and what could be the possible mechanisms of detecting it. 

Following Refs.~\cite{Bauer:2017qwy,Schmeier:2013kda,Agrawal:2010fh}, our work assumes that the interaction of any Weakly Interactive Massive Particle (WIMP) with the SM is mediated by a new massive gauge boson, which we will indicate hereafter by $\zp$. This $\zp$ boson then acts as a mediator in the production of primordial DM until the \textit{freeze-out} is reached \cite{Kolb:1990vq}. A higher mass mediator is preferred due to strong experimental constraints in the search for a resonance at lower masses, so we will show that a massive vector mediator on the TeV scale would be accessible even with the restrictions imposed on phase space by the current collider searches. Unlike the cases analyzed in Refs.~\cite{Schmeier:2013kda,Agrawal:2010fh}, we do not assume, \textit{a priori}, any effective model and proceed with the calculation of the total cross section, $\xstot$, using the Feynman rules obtained from the model Lagrangian. Furthermore, in all the processes described in this work, DM is elastically scattered, but inelastic or scatterings described with form factors could also be analyzed \cite{Feldstein:2009tr,Cui:2009xq}. Hence, we are not only interested in parameters related to the final state of the particles, such as their mass and spin, but also assume characteristics for a mediator $\zp$ and their couplings with the initial state fermions, $\psi$, and the DM particle, $\chi$, in final state. By including a mediator in the process, $\psi\bar\psi\to\zp\to\chi\bar\chi$, we increase the number of parameters involved and, thus, evaluate a better theoretical agreement of the model with the current experimental data.

In this work we explore the $\zp$ production in electron-positron annihilation ($\ee$) at CLIC at $\sqrt{s}=$~3~TeV and for proton-proton ($pp$) collisions at the LHC at $\sqrt{s}=$~14~TeV. Considering the limits already imposed by the searches performed by the ATLAS and the CMS Collaboration of DM mediator mass above 2~TeV, we focus this study in a $\zp$ mass of 3~TeV in both CLIC and LHC. This choice allow us to investigate the DM production near the $\zp$ resonance where the event rate should be enhanced by the resonance cross section. As a result, one has to properly account for the viability of the DM candidate by accounting for the expected relic density, and this calculation cannot rely on the usual framework away from resonances. As shown in Ref.~\cite{Gondolo:1990dk, Griest1991}, the proper treatment of the relic density near a resonance results in a weaker reduction of the expected relic density. Hence, we compare our predictions with the proper evaluation of the relic density for DM species.

If DM was produced by a thermal process on which a freeze-out has occurred \cite{Kolb:1990vq}, then we could attempt to recreate it with the use of particle colliders with sufficient high energies. An arbitrary coupling of DM with ordinary matter is hence assumed, expressed in the form of a gauge coupling $g_{\chi}$, which could represent a direct DM coupling to leptons and quarks \cite{Bauer:2017qwy} or interposed by a massive mediator \cite{Langacker:2008yv,Frank:2019nwk,Buchmueller:2014yoa}. Nonetheless, the detection of DM particles poses a major experimental challenge, since DM-related couplings are expected to be very weak \cite{Boveia:2018yeb}, e.g., as much or even more than those with neutrinos, and exclusion limits have been recently imposed on massive vector mediators up to the TeV scale \cite{CMS:2017zts,CMS:2018mgb,ATLAS:2017bfj}. In general, searches in high energy colliders focus in the observation of DM signatures in the form of missing transverse momentum or missing transverse energy \cite{Linssen:2012hp,Kadota:2018lrt,Fox:2011pm}. Such a signal would occur if the DM particles are invisible to the detector or a possible charged DM particle has a sufficiently long lifetime to pass through the detector volume and leave a characteristic trace of charged particles, decaying shortly thereafter into particles too light to generate any signs on the detector calorimeters. Alternatively, the production of DM may be detected by the emission of SM particles in the initial interaction state, where only SM particles are involved. Searches known as \textit{mono-X} may indicate the associated production of jets, vector bosons ($H$,$Z$), photons, among others. For instance, the future experiments at CLIC will be especially sensitive to wide searches of DM in mono-photon production \cite{Blaising:2021vhh}. Hence, we investigate the DM production via massive vector mediator in mono-photon processes at high-energy colliders.

This paper is organized as follows: in \autoref{modeling} we describe the theoretical modeling of the $\zp$ mediator, with the respective motivations for a scalar and fermion final DM states. Here we assume that the DM in the final state composes all the DM relic abundance observed by the PLANCK satellite \cite{Planck:2018vyg}, nonetheless a richer dark sector can be studied in the TeV scale in a subsequent analysis, as proposed by \cite{Oncala:2021tkz}. Section~\ref{reliccalc} presents the calculation of the relic density at the resonance for the comparison with our results in high-energy colliders to evaluate the regions where DM production is still possible. In \autoref{results} we discuss how this model could be perceived with a ISR assuming a hard-photon emitted by the incoming fermions \cite{Bonneau:1971mk} and present the results obtained for the mass and coupling constant regions available at the high-energy CLIC and LHC colliders. Finally, we present our conclusions in \autoref{conclusions}.

\section{Theoretical framework}
\label{modeling}

Extensions of the SM can usually be studied using effective, simplified, or (so to say) complete models \cite{Bauer:2017qwy,ParticleDataGroup:2022pth}. Still, we can make several claims regarding the nature of the kind of New Physics we expect to find even with the simplest effective models \cite{Wudka:1996ah}. These models are a starting point for studying New Physics, given their simplicity on describing the particles and interactions involved using only a small number of parameters that can be directly related to experimental observations, such as: mass of the particles involved, their decay widths, production cross sections of these new processes, among others \cite{LHCNewPhysicsWorkingGroup:2011mji}. One can apply a simplified model in trying to explain some New Physics results through functions of the variables involved in its description, excluding certain values based on different experimental constraints \cite{Abercrombie:2015wmb}.

We start with interaction Lagrangians describing a SM extension with a new renormalizable symmetry group $U_{\chi}(1)$ acting as a vector portal for DM. The use of a $U_{\chi}(1)$ symmetry for investigating interactions between DM and SM has been widely proposed \cite{Roszkowski:2017nbc,ParticleDataGroup:2022pth,Kopecky:2012oda,Langacker:2008yv} and very tightly constrained at low and high masses have been imposed mostly by collider experiments (e.g., see Refs.~\cite{ATLAS:2021kxv,CMS:2021far}).  This work explores a framework to probe the limits and to analyze parameters for the DM thermal production through a process that involves interactions of the SM with the dark sector mediated by a new massive boson mediator ($\zp$) described with a Breit-Wigner (BW) resonance. Such mediator couples to scalar and fermion fields as candidates for DM. Feynman diagrams representing the $s$-channel $\zmp$ exchange with a DM candidates, $\chi$, are shown in \autoref{SFV_diagram}, where $\grl$ and $\gx$ are the couplings of this vector boson to the SM and the different DM fields, respectively. Here we treat two different possibilities for the final state; details of the implications of the simultaneous existence of these final states for DM and possible interactions between them are beyond the scope of this work. These possible states would be an aspect of a even more complete model where further studies could be performed with a experimental observation. We then focus on the evaluation of the cross section and in turn analyze the parameter space more comprehensively. 

\subsection{Tree-level process with s-channel resonance}

\begin{figure}[!t]
\centering
\includegraphics[width=0.75\textwidth]{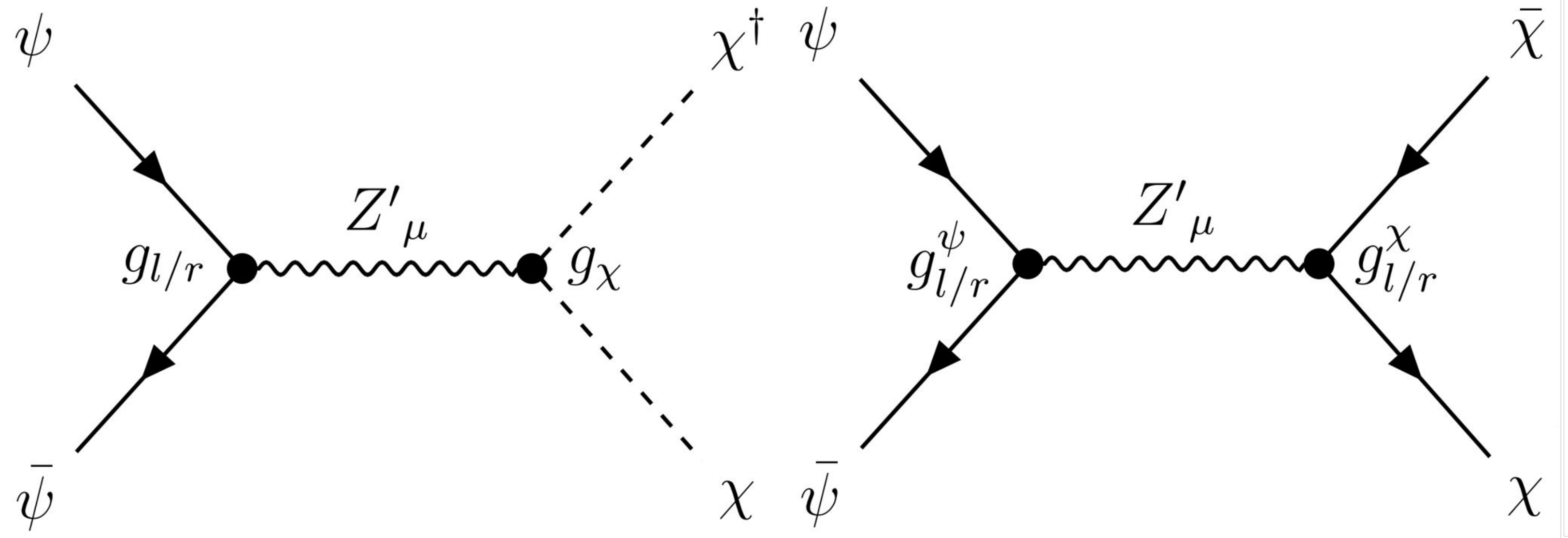}\caption{Feynman diagrams for the interaction of SM fermions, $\psi$, with a scalar (left), fermion (center), and vector (right) DM, $\chi$, field through a $\zp$ boson. The couplings $\grl$ and $\gx$ represents the coupling of the $\zp$ boson with the SM and the DM fields, respectively.} \label{SFV_diagram}
\end{figure}

Let $\psi$ be any SM fermion spinor and $\zmp$ a real vector field corresponding to an on-shell massive vector boson coupling scalar particles representing the DM fields [\autoref{SFV_diagram} (left)]. An interaction Lagrangian of this $s$-channel can be written as 
\begin{equation}\label{lescalar}
\mathcal{L}_{\rm{int}}^{\rm{scalar}} = -\frac{1}{4}F^{\mu\nu}F_{\mu\nu} + \frac{1}{2}\Mdf^2{\zpp}{\zmp} + \bar \psi\gamma^\mu\left(\gl P_L + \gr P_R\right)\psi \zmp+\gx\left(\chi^\dag\partial_\mu\chi - \chi\partial_\mu\chi^\dag\right){\zpp},
\end{equation}
where we use $\Mdf$ for the mediator mass and $ \gamma^{\mu} $ are the usual Dirac matrices. Besides that, $\zmp$ indicates a real vector field with field tensor given by $F_{\mu\nu}\equiv\partial_\mu{\zp_\nu}-{\partial_\nu}\zmp$, which, along with the mass element defined by the term $\frac{1}{2}\Mdf^2{\zpp}{\zmp}$, composes the kinetic term of the $\zp$ boson. The $P_L$ and $P_R$ operators refer to left and right-handed operators, respectively, defined by $P_L \equiv\frac{1}{2}(1-\gamma^5)$ and $P_R\equiv\frac{1}{2}(1+\gamma^5)$, with $\gl$ and $\gr$ representing chiral coupling magnitudes. The final DM scalar state is well motivated both in simplified effective models and more complete models containing sometimes a Higgs doublet \cite{Chang:2017gla,Arcadi:2019lka,Ruhdorfer:2019utl} that can act as a mediator between or be the main composition of a dark sector. In most of the literature, masses below a few GeV are largely excluded by different experimental DM detection pathways \cite{XENON:2017lvq,XENON:2022ltv,Falkowski:2018dsl}, hence we discuss the production of scalar DM where the final state mass is on the TeV scale.

For a DM particle characterized as a Majorana fermion [\autoref{SFV_diagram} (center)], the interaction Lagrangian has the form
\begin{equation}\label{lfermion}
\mathcal{L_{\rm{int}}^{\rm{fermion}}} = -\frac{1}{4}F^{\mu\nu}F_{\mu\nu} + \frac{1}{2}\Mdf^2{\zpp}{\zmp} + \left[ \bar\psi{\gamma^\mu}\left(\gl P_L + \gr  P_R \right)\psi + \bar\chi{\gamma^\mu }\left(g_{\chi} P_L + g_{\chi} P_R \right)\chi  \right]{\zmp},
\end{equation}
where we adopt $ g_{l/r} \neq g_{\chi}$ and $g_{\chi , r} = g_{\chi , l} = g_{\chi}$ due to the Majorana condition for the spinor components. Moreover, we see that the interaction of the $\zp$ boson with the DM field is given by the operators in the last term of \autoref{lfermion}, which also contains the adjoint spinor for the DM particle $\bar\chi$ and once again the chiral operators and couplings. Fermion DM is well motivated through the literature and seen as one of the main candidates for WIMP DM in many different models \cite{Roszkowski:2017nbc,ParticleDataGroup:2022pth,Bertone:2004pz,CMS:2021far}. For instance, a well-studied case in minimal supersymmetric models (MSSM) \cite{Cabrera:2016wwr} would be the existence of a long lived particle (LLP) in the form of a neutralino, the Dirac or Majorana fermion coming from the symmetry of the neutral mediators of the SM. Furthermore, universal extra-dimensional models \cite{Cornell:2014jza} as well as models with sterile neutrino \cite{Boyarsky:2018tvu} introduce candidates in the form of a fermion final state, sometimes discussed in another mass scale though. We emphasize that we deal with masses at the TeV scale, an accessible mass window in searches at high-energy collider experiments.

\subsection{Production cross section and decay widths}

The cross sections for the three process in \autoref{SFV_diagram} are obtained with the help of \texttt{FeynCalc}~\cite{Shtabovenko:2020gxv} and \texttt{FeynArts}~\cite{Hahn:2000kx} packages available for the \texttt{Wolfram Mathematica} software \cite{Mathematica}. From the Lagrangians and Feynman diagrams, scattering amplitudes are obtained and evaluated using Feynman rules and appropriate kinematic variables for these packages. Once the expression for the total cross section of the process $2 \to 2$ is obtained, the mass and couplings of the particles involved in the process are treated as free parameters and evaluated separately.

The total cross section, $\xstot$, for a process $2 \to 2 $ in the center-of-momentum (CM) frame can be calculated in terms of the Mandelstam variables,
\begin{equation}\label{goldenrulet}
\frac{\dif\xs}{\dif t} = \frac{1}{16\pi }\frac{1}{{s(s - 4m_\psi^2)}}\left|{\mathcal{ M}\left({s,t}\right)}\right|^2.
\end{equation}
where $s$ and $t$ are the squared of momenta in $s$ and $t$ channels and $m_\psi$ is the SM fermion mass and we take the limit with massless initial fermions. Here we perform the average over the spin of the initial states of those processes and the square of weighted scattering amplitude over all initial spin states. Hence, we employ the CM framework to analyze the available phase space based on the possible couplings of $\zmp$ and the DM particles. Finally, the cross section for these processes result in: 
\begin{subequations}\label{totalXsec}
\begin{equation}\label{totalXsec-scalar}
\hat\xs^{\rm{scalar}} = \frac{\gx^2 \left(\gl^2+\gr^2\right) \left[s(s-4\mx^2)\right]^{3/2}}{192\pi s^2\left[ \left(s-\Mdf^2\right)^2 + \Gamma^2\Mdf^2\right]},
\end{equation}
\begin{equation}\label{totalXsec-fermion}
\hat\xs^{\rm{fermion}} = \frac{\gx^2 \sqrt{s-4\mx^2} \left[\gl^2\left(s-\mx^2\right) + 6\gl\gr \mx^2+\gr^2\left(s-\mx^2\right)\right]}{48 \pi  \sqrt{s}\left[\left(s-\Mdf^2\right)^2 + \Gamma^2 \Mdf^2\right]}.
\end{equation}
\end{subequations}
The terms for the BW width ($\Gamma^2\Mdf^2$) in the denominator of the scattering amplitudes correspond to the mediator exchange of a $s$-channel 
resonance.

As we will deal with the cross section of a process involving a massive vector mediator, we need to compute the decay widths. The $\zp$ decay into both species of DM particles are evaluated to determine the decay width, $\Gamma^{\rm{i}}$, at which it can decay into two DM particles of identical masses $\mx$. The calculation is performed in the same way as for any $1\to 2+3$ process, where the decay width $\Gamma$ takes the form:
\begin{equation}\label{goldenruledecay}
\Gamma_{a\to b+c} = \frac{\left|\vec{p}_f\right|}{8\pi M^2}\frac{{\left|\mathcal{M}_{1\to 2}\right|}^2}{3},
\end{equation}
resulting in the respective decays widths for each DM final state:
\begin{equation}\label{decay-scalar}
\Gamma^{\rm{scalar}} = \frac{\gx^2 \left(\Mdf^2-4 \mx^2\right) \sqrt{1-4\mx^2/\Mdf^2}}{48\pi\Mdf},
\end{equation}
\begin{equation}\label{decay-fermion}
\Gamma^{\rm{fermion}} = \frac{\left[\glx^2 \left(\Mdf^2-\mx^2\right) + 6\glx\grx\mx^2+\grx^2 \left(\Mdf^2-\mx^2\right)\right]\sqrt{1-4 \mx^2/\Mdf^2}}{24\pi\Mdf}.
\end{equation}
We can separate the SM coupling in three different types due to the nature of the mediator. From left and right projection operators, the $\zp$ mediator can be vector, axial-vector, or chiral, with coupling constants according to \autoref{couplings}, where we use $\glx=\grx=1$ in the fermion case \cite{Schmeier:2013kda}.

\subsection{Initial-state photon radiation}

The experimental detection of DM is a hard task given the unknown characteristics of its interaction with ordinary matter, hence ways of detecting it is a topic of intense research \cite{Cebrian:2022brv}, typically leading to the search of events with missing transverse energy (MET), $\MET$. One way of observing a DM event with MET is to consider the emission of SM particles as initial state radiation (ISR). As such, different particles can be emitted from the initial colliding particles, collectively known as mono-$X$ searches. The search for a photon as ISR is natural as electromagnetic radiation is perhaps one of the simplest to be measured with precision at experiments in particle colliders. Also, it can have a very broad spectrum, on scales from keV to TeV, depending on the invisible event that one want to characterize \cite{Kalinowski:2021vgb,Kalinowski:2022cnt,Kopecky:2012oda,Tolley:2016lbg}. 

\begin{figure}[!t]
\centering
\includegraphics[width=0.5\textwidth]{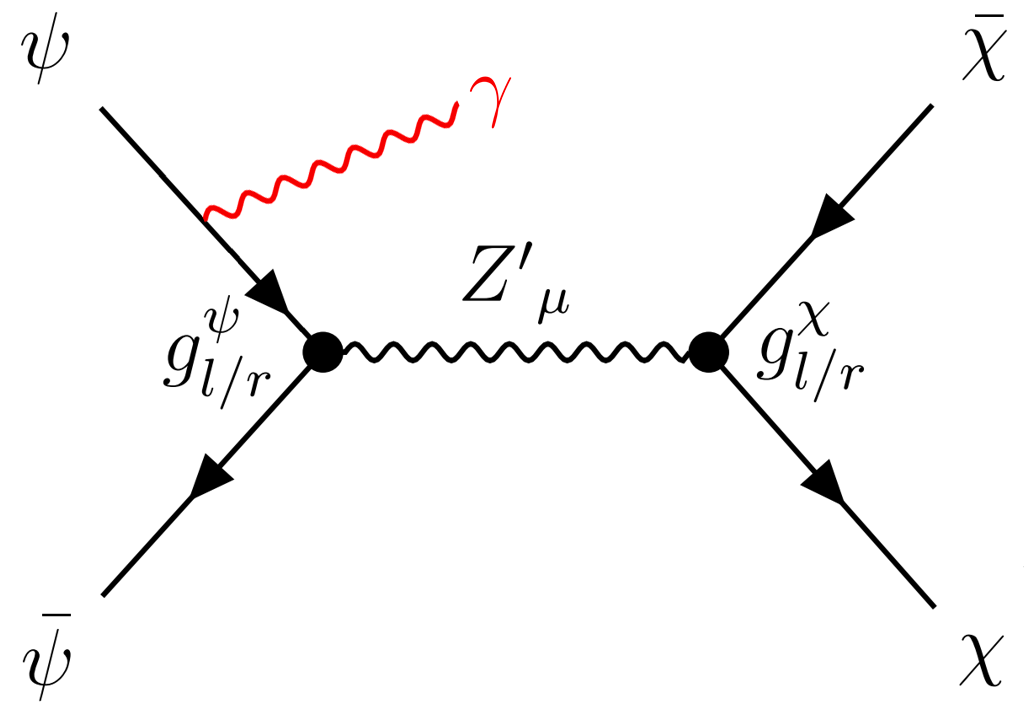} 
\caption{A representation of a Feynman diagram with the emission of a photon as ISR, in red wiggly line, in the case of fermion DM production.}
\label{HF_diagram}
\end{figure}

\begin{figure}[!t]
\includegraphics[width=0.75\textwidth]{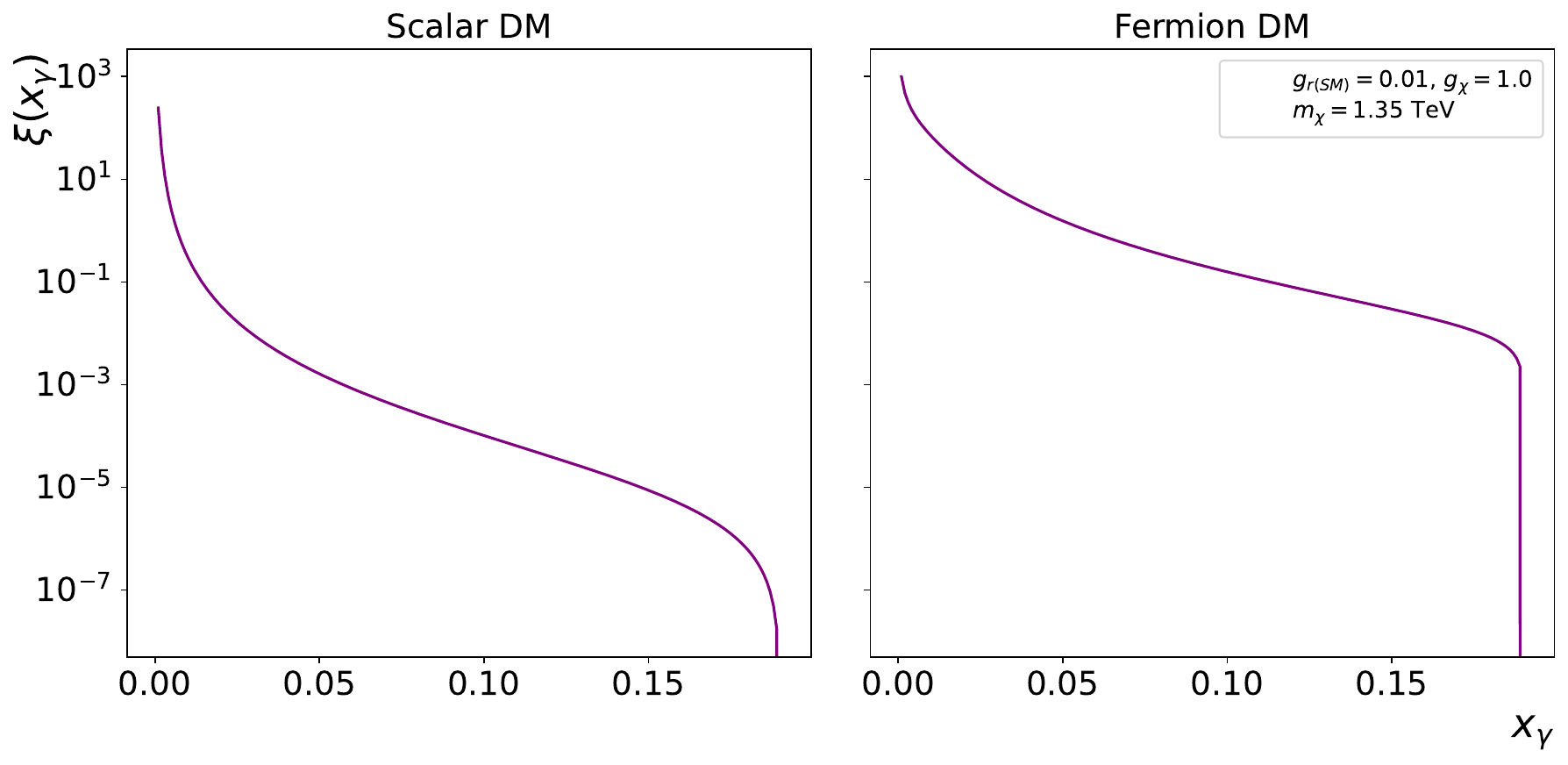}
\includegraphics[width=0.75\textwidth]{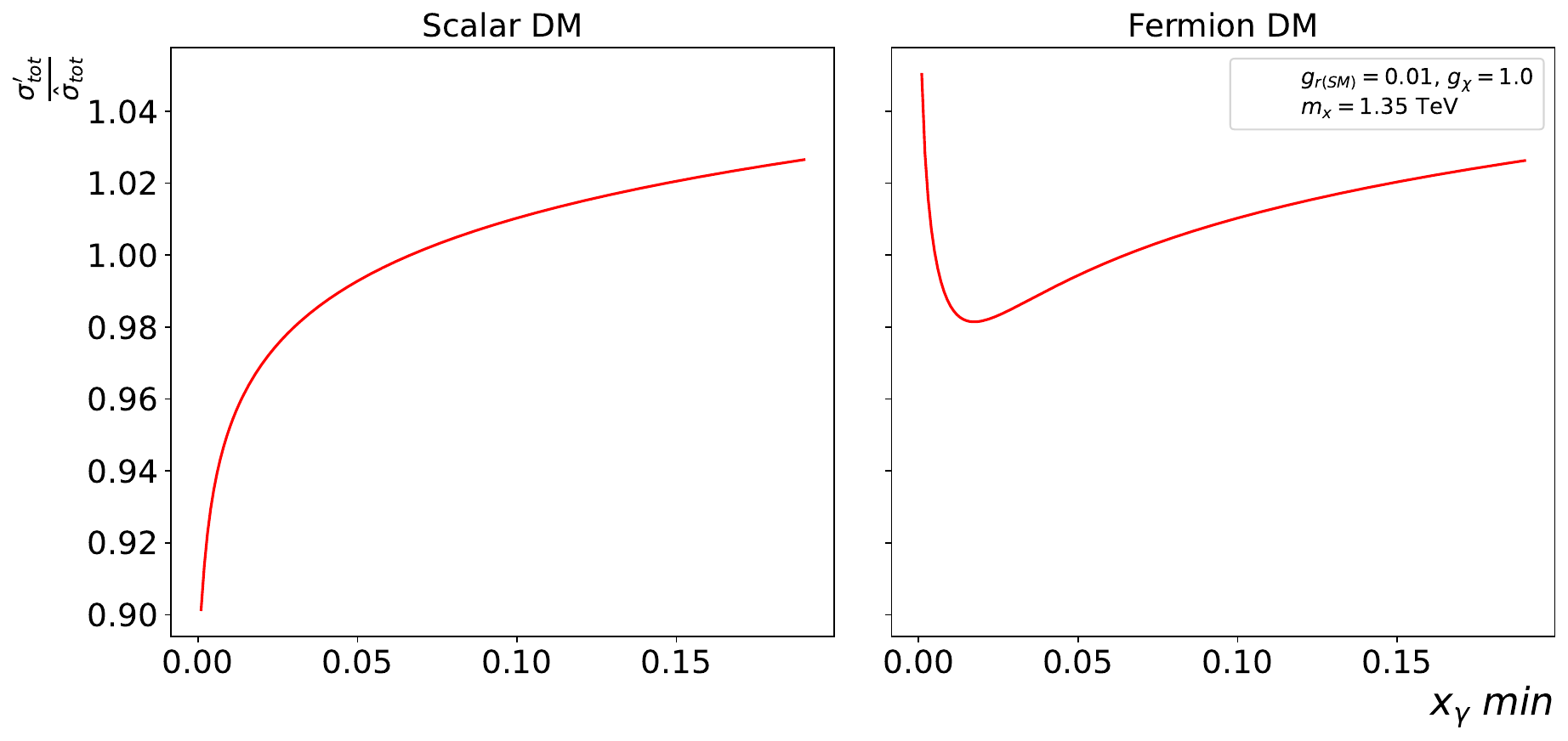}
\caption{\label{xiratio}Top: correction $\xi(\xg)$ in terms of the photon energy fraction, which account for radiative corrections. The slope shows the different contributions of such corrections, increasing $\delta$ at very low photon energies and rapidly decreasing for more energetic photons. Bottom: the ratio of the cross section with ISR, $\xstotl$, and without ISR, $\hxstot$ that represents the value $1+\delta$ carrying all radiative corrections. In both cases the cross sections are computed with $\Mdf=3$~TeV and $\mx=$~1~GeV.}
\end{figure}

Radiative corrections to the tree-level DM process are necessary to account for DM production via mono-photon process. While most of the studies in the literature are given in terms of an approximation with a soft photon or calculated numerically for a $3 \to 2$ process, we choose here to account for the radiative correction with emission of a hard photon as discussed in Ref.~\cite{Bonneau:1971mk}. This framework proposes a factorization in terms of lower order processes where $\psi\bar{\psi}$ reactions generate a final state with the emission of a hard photon in the event. This particular approach uses a rigorous calculation of higher order corrections together with a more precise evaluation of the phase space of the emitted photon. Such precautions are necessary due to the high energy involved and the emitted photon itself, which would disqualify a process containing only one ISR of a low energy or soft collinear photon, as in the Weizsäcker-Williams (WW) approximation ~\cite{Schmeier:2013kda,Chen:1975sh}. We employ this framework by using the appropriate factorization to obtain the total cross section for a mono-photon process, as represented in \autoref{HF_diagram}:
\begin{eqnarray}\label{xsecgamma}
\xstotl(\psi\bar\psi\to\zp\gamma\to\gamma\chi\bar\chi)=\hxstot(\psi\bar\psi\to\zp\to\chi\bar\chi)(1+\delta),
\end{eqnarray}
with $\hat\sigma$ given by \autoref{totalXsec} integrated over the scattering amplitude squared for each DM species and
\begin{eqnarray}\label{ISRequation}
\delta &=& \frac{2\alpha}{\pi}\left\{\left( -1+2\log\frac{\sqrt{\hat{s}}}{m_\psi}\right)\left[ {\log{\xgmin}} + \frac{13}{12} + \int_{\xgmin}^{\xg^{\max}}\dif \xg ~ \xi(\xg) \right] - \frac{{17}}{{36}} + \frac{\pi^2}{6} \right\},\\
\xi(\xg) &=& \frac{1}{x_\gamma}\left(1-\xg+\frac{\xg^2}{2}\right) \frac{\hat\sigma (\hat{s}-\hat{s}{\xg})}{\hat\sigma(\hat{s})},
\end{eqnarray}
where $\alpha=1/137$ is the electromagnetic fine structure constant, $q_\gamma$ is the momentum carried by the photon from ISR, and $\hat\sigma(\sqrt{\hat{s}})$ is the total cross section of the $2 \to 2$ process, with energy $\sqrt{\hat{s}}$. Function $\xi(\xg)$ takes into account the available center-of-mass energy for producing the DM pair plus a hard photon, which decreases for more energetic photons. This can be seen in \autoref{xiratio} (top panel) where the function $\xi(\xg)$ is shown, which depends on the partonic cross section with and without ISR, for both DM species. The shape of the distribution is very similar in all three cases, although the fermion DM a little enhancement in the tail towards high $\xg$. A typical photon spectrum in particle detectors starts at a few GeV where isolation and reconstruction efficiencies are above 90\%.

Considering a usual efficiency turn-on curve in trigger selection, we employ $\xg=q_{\gamma}/E_{beam}$ with a minimum photon energy of 60~GeV, or $\xg=0.04$ for a beam of 1.5~TeV, motivated by the identification capabilities expected at CLIC dp \cite{Blaising:2021vhh} and assessed in the LHC experiments \cite{CMS:2015myp}. \autoref{xiratio} (bottom panel) shows the distribution of the ratio between the total cross sections with and without ISR in terms of the minimum photon energy $\xgmin$, which illustrate the contribution of the $\delta$ term in \autoref{xsecgamma}. Although the shape of $\xi(\xg)$ decreases rapidly with increasing photon energy fraction, the radiative corrections improve the production cross section at higher photon energies up to $\sim$7\%, which is independent of the dark sector parameters, such as masses and coupling constants. Also, the dip structure occurring around $\xgmin\approx0.02$ reveals the effect of the corrections for very low photon energies. The ratio achieves a value of 1 at $\xgmin\approx0.06$, starting point where the radiative corrections increase the total cross section. Although the chosen value of 0.04 as minimum photon energy in this study is located in a region where the cross section is slightly suppressed, we aim to investigate the kinematic region expected for the photon identification in experiments, especially at CLIC dp.

As shown in Ref.~\cite{Kalinowski:2020lhp}, the production cross section of DM pairs plus hard photon are ${\cal{O}}$(fb), making its observation possible with an integrated luminosity of the order of ab$^{-1}$. Similar simulations also indicate high visibility in the emission and consequent detection of mono-photons from invisible decays, with significant cross section and transverse momentum fraction of the emitted photon \cite{Kalinowski:2022cnt}. The detection of such mono-photon event can be made with the usual signature of high-$p_T$ photon plus MET, where MET will peak around the resonance mass of the mediator.

\section{Calculation of the DM relic density near a resonance}
\label{reliccalc}

The DM abundance and, consequently, the cross sections involved in its primordial production, are typically calculated taking into account processes in equilibrium and away from poles or production resonances of a given species. When considering a resonant production processes, like the one investigated in this work, one cannot simply apply the usual solution of the Boltzmann equation in terms of $\sigmav$. As described in Ref.~\cite{Gondolo:1990dk}, one can write $\sigmav$ as a non-relativistic BW resonance (like it is the case for a cold DM candidate in the LCDM model) in the following form
\begin{equation}\label{sigmavres}
\sigmav_{\text{res}} = \frac{16\pi}{\mx^2}\frac{(2J+1)}{(2S+1)^2}x^{3/2}\pi^{1/2} ~ \frac{\Mdf\Gamma_{\zp}}{\mx^2}B_i(1-B_i) \,F_{l}(z_R;x),
\end{equation}
where $x=\mx/T$ and
\begin{equation}
F_{l}(z_R;x) = \text{Re}\frac{i}{\pi}\int_{0}^{\infty} \frac{(1+\epsilon)^{1/2} e^{-x\epsilon}}{(1+2\epsilon)(z_R-\epsilon)}\dif\epsilon.
\end{equation}
The quantities $J$ and $B_i$ are, respectively, the spin and resonance branching fraction of the initial state\footnote{Note that we consider the annihilation of DM particles into SM ones via the resonant mediator $\zp$ in the relic density calculation, that is, in the cosmological context of this equation, the initial state particles are the DM candidates themselves.}, whereas $S$ and $\mx$ are the spin and mass of the DM particle, respectively. We also introduce the energy per unit mass of the process as a whole,
\begin{equation}
\epsilon=\frac{s-\mx^2}{\mx^2},
\end{equation} 
and an auxiliary variable in terms of the masses and decay widths of the particles involved
\begin{equation}
z_R=\frac{\Mdf^2-\mx^2}{\mx^2}+i\frac{\Mdf\Gamma_{\zp}}{\mx^2}.
\end{equation}
This expression allows us to numerically estimate the dimensionless density parameter referring to the primordial DM fraction as \cite{Gondolo:1990dk,Arcadi:2017kky}
\begin{equation}\label{omega-relic}
\Omega_{\text{DM}}h^2 \approx 8.76\times 10^{-11}\,\mbox{GeV}^{-2}\left[\int_{T_0}^{T_f} g_{*}^{1/2}\langle\xs v\rangle_\text{res}\frac{\dif T}{\mx}\right]^{-1},
\end{equation}
being $T_0$ and $T_f$ the current and at \textit{freeze-out} temperatures, respectively, and $g_{*}$ are the particles degrees of freedom in the same epoch.

\begin{figure}[!t]
\centering
\includegraphics[width=1\textwidth]{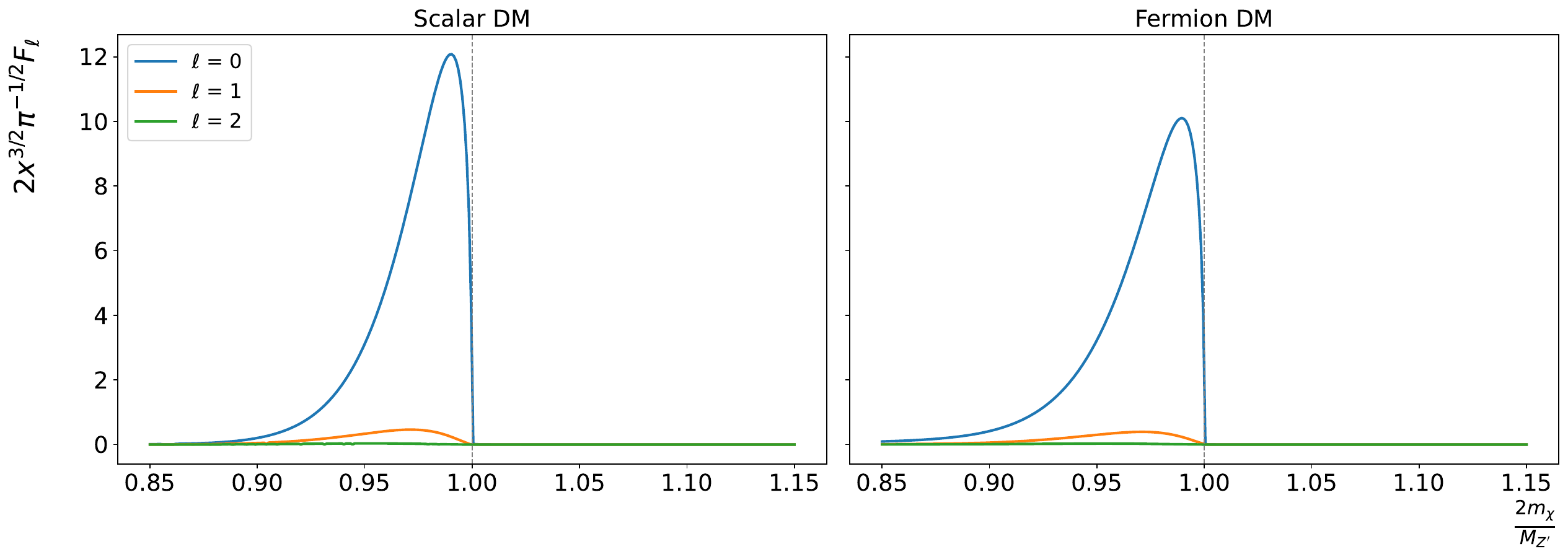}
\caption{\label{fig-profile}Profile of the thermal average, $\Sigma(z_R;x)$, near the $\zp$ resonance for $l=0,1,2$.}
\end{figure}

We can also check the profile of the thermal average near the resonance computed in the proper framework for the expansion of the Breit-Wigner cross section. As we can see in \autoref{fig-profile}, the profile defined as
\begin{equation}
\Sigma(z_R;x) = 2x^{3/2}\pi^{-1/2}F_{l}(z_R;x)
\end{equation}
shows the relative masses where the enhancement in the thermal average occurs near the $\zp$ resonance, for relative masses above 0.8. For relative masses below threshold the thermal average is very similar if compared to the usual, non-resonant calculation of the thermal average. The enhancement peak is positioned at DM masses below the resonance mass given that fact that there is enough thermal energy to produce a heavier resonance during the equilibrium phase. Also, the profile does not extend beyond the resonance mass since we are taking into account a narrow resonances for the $\zp$ boson for both scalar and fermion DM (both $\Gamma_{\zp}/\Mdf\sim 10^{-2}$). DM particle with masses above the resonance mass would be possible for wider resonances, where low energy tail would allow a non-zero thermal average for DM particles with relative mass above the resonance mass. As a result, we consider in this study $\mx$ masses with relative mass in the range $0.8 < 2\mx/\Mdf < 1.0$ in order to probe the region where the thermal average is enhanced by the $\zp$ resonance: together with the relic density, results will contain a (red dashed) line delimiting the threshold relative to $2\mx/\Mdf=0.8$.

Taking for instance a resonance of 3~TeV, the threshold of $\mx$ with enough thermal energy to produce a $\zp$ starts at 1350 GeV -- we use this mass value as reference. Furthermore, the production process with ISR photon needs to probe the enhanced region above $\sqrt{s}=0.8\Mdf$. Thus, the reduced beam energy resulting from the photon emission cannot go below this threshold if we want to investigate the mass region near the $\zp$ resonance, restricting the photon energy for possible searches of DM production. In this work we consider a photon irradiation from an incoming fermion with energies ranging from 60~GeV up to $0.2\Mdf$, allowing to evaluate the production cross section where the relic density will be lesser suppressed near the resonance.

\begin{figure}[!t]
\centering
\includegraphics[width=1\textwidth]{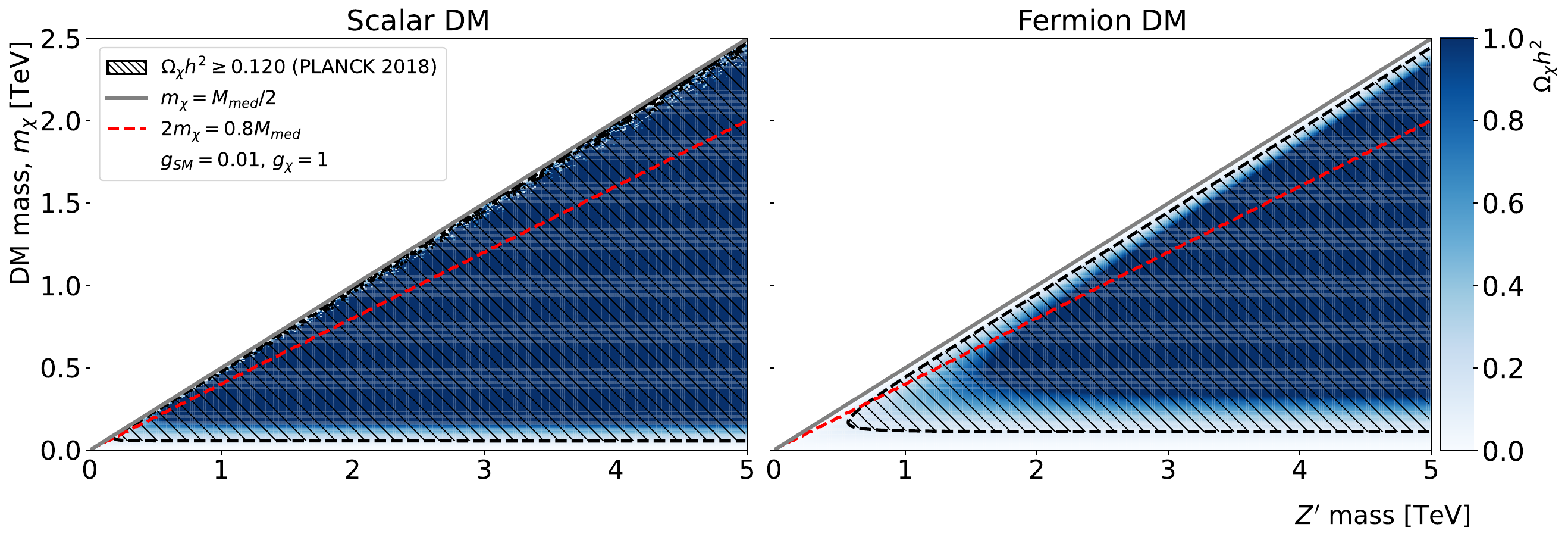}
\caption{\label{omega-relic-fig_ee}Dimensionless density parameter $\Omega_\chi$ for DM produced by a resonance with mass $\Mdf$ ($\zp$ mass) and DM pairs with mass $\mx$ taken for $\ee$ collisions at $\sqrt{s}=$~3~TeV. The hatched area shows the DM overabundance than the observed CDM abundance, which is excluded. The diagonal solid line represents the kinematic limit with $\mx=\Mdf/2$. For this plot, $\gr=0.01$ and $\gx=1$. The red dashed line shows the threshold of the $\mx$ to produce the resonance with thermal energy.}
\end{figure}

\begin{figure}[!t]
\centering
\includegraphics[width=1\textwidth]{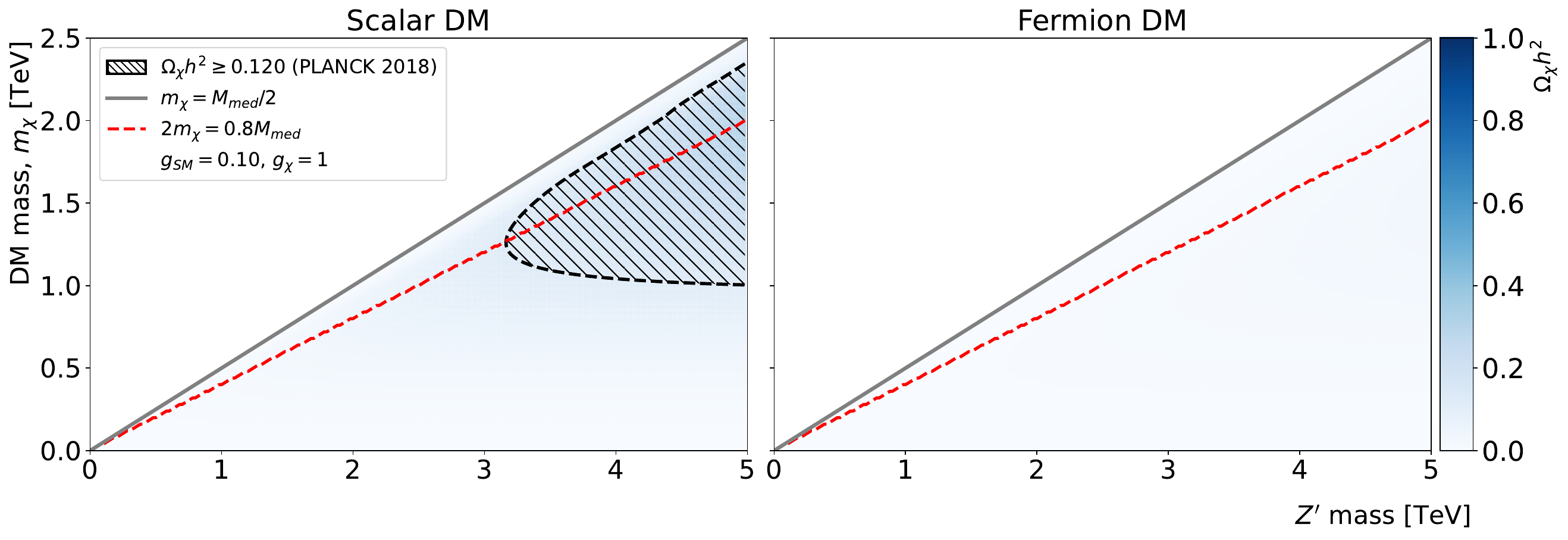}
\caption{\label{omega-relic-fig_qq} Same as \autoref{omega-relic-fig_ee}, but for $pp$ collisions at $\sqrt{s}=$~14~TeV and with $\gr=0.10$.}
\end{figure}

Following recommendations in the literature \cite{CMS:2021far,CMS:2021dzg}, the coupling of the mediator with the SM is fixed: $\grl = 0.01$ for the $\ell\ell\zp$ coupling in $\ee$ collisions and $\grl = 0.10$ for the $q\bar{q}\zp$ coupling in $pp$ collisions, whereas the coupling of $\zp$ with the dark sector is set at $\gx=1$. Figures~\ref{omega-relic-fig_ee} and \ref{omega-relic-fig_qq} show how the DM overabundance regions, behave in the mass scans $\mx\times\Mdf$, parameterized by the dimensionless density $\Omega_\chi$ obtained with \autoref{omega-relic}. The region that comprises the limit with $\Omega_{\chi}h^2\geq 0.120$ corresponds to the hatched area and is used in all results indicating the region of excess primordial DM production. We note that the regions in the mass scan differ significantly from those shown in other works that do not take into account processes near or at the resonance peak of a massive mediator, even though numerically calculated for analogous processes \cite{ATLAS:2021kxv,CMS:2021far,Albert:2017onk}. These results show that the proper calculation of these limits is essential to evaluate the available regions to probe the masses of the mediator and the DM particles taking into account the limits obtained from astrophysical observations. While all mass region is mostly excluded for CLIC energy regime (\autoref{omega-relic-fig_ee}), the results are more significant for the LHC kinematics, where much small region of the phase space are excluded for scalar DM and completely accessible for fermion DM.

\begin{figure}[!]
\includegraphics[width=\textwidth]{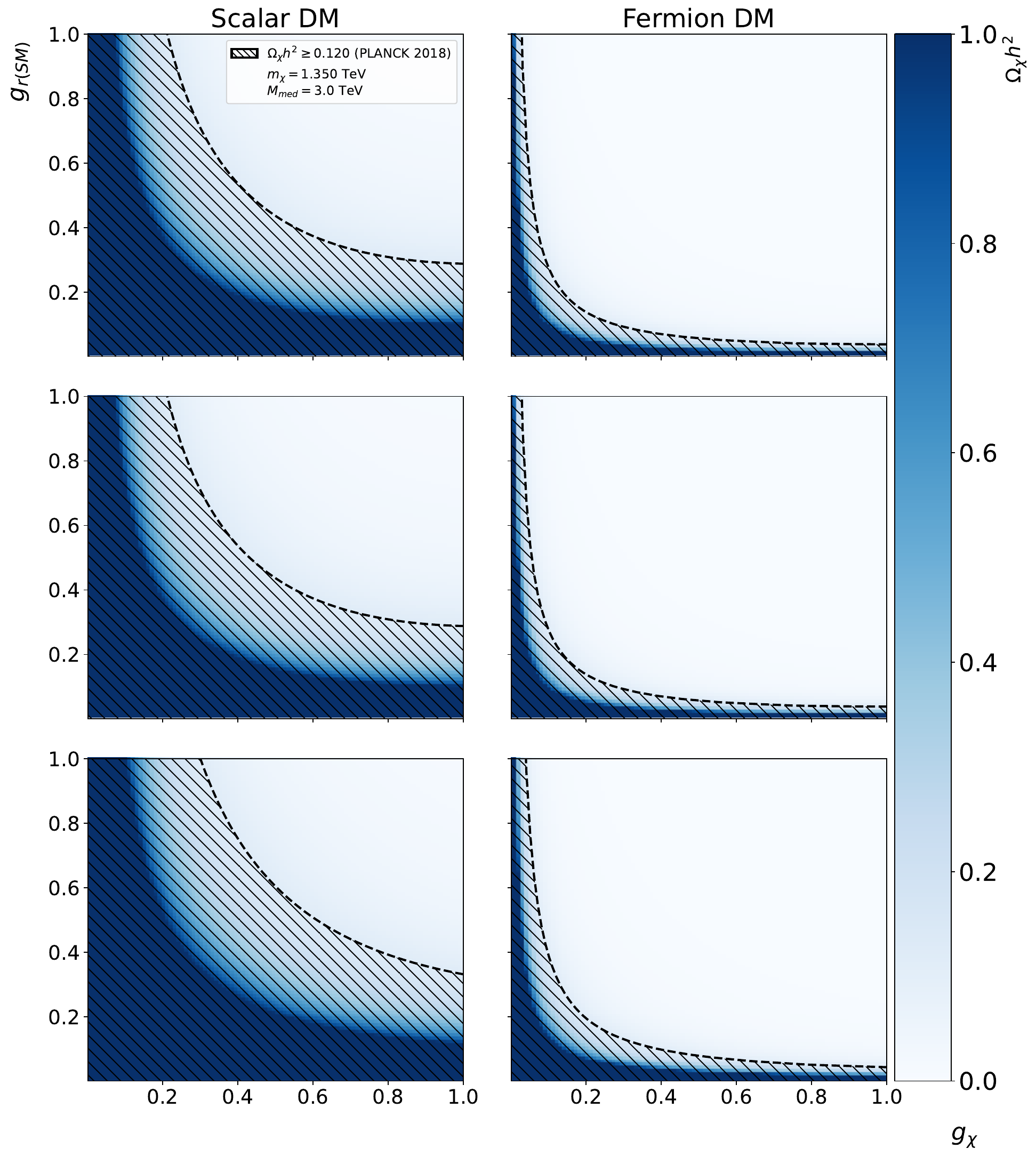}
\caption{\label{omega-relic_coup_ee}Dimensionless density parameter $\Omega_\chi$ for DM produced by a resonance with mass $\Mdf=3$~TeV and decaying into DM pairs of mass 1350~GeV in $\ee$ collisions at 3~TeV. The values shown here were obtained using \autoref{omega-relic}. The region with a cross section smaller than that needed to produce the observed CDM abundance is shown beveled in black. We have vector couplings in the first row, axial-vector couplings in the second, and chiral (right) in the third one.}
\end{figure}

\begin{figure}[!]
\includegraphics[width=\textwidth]{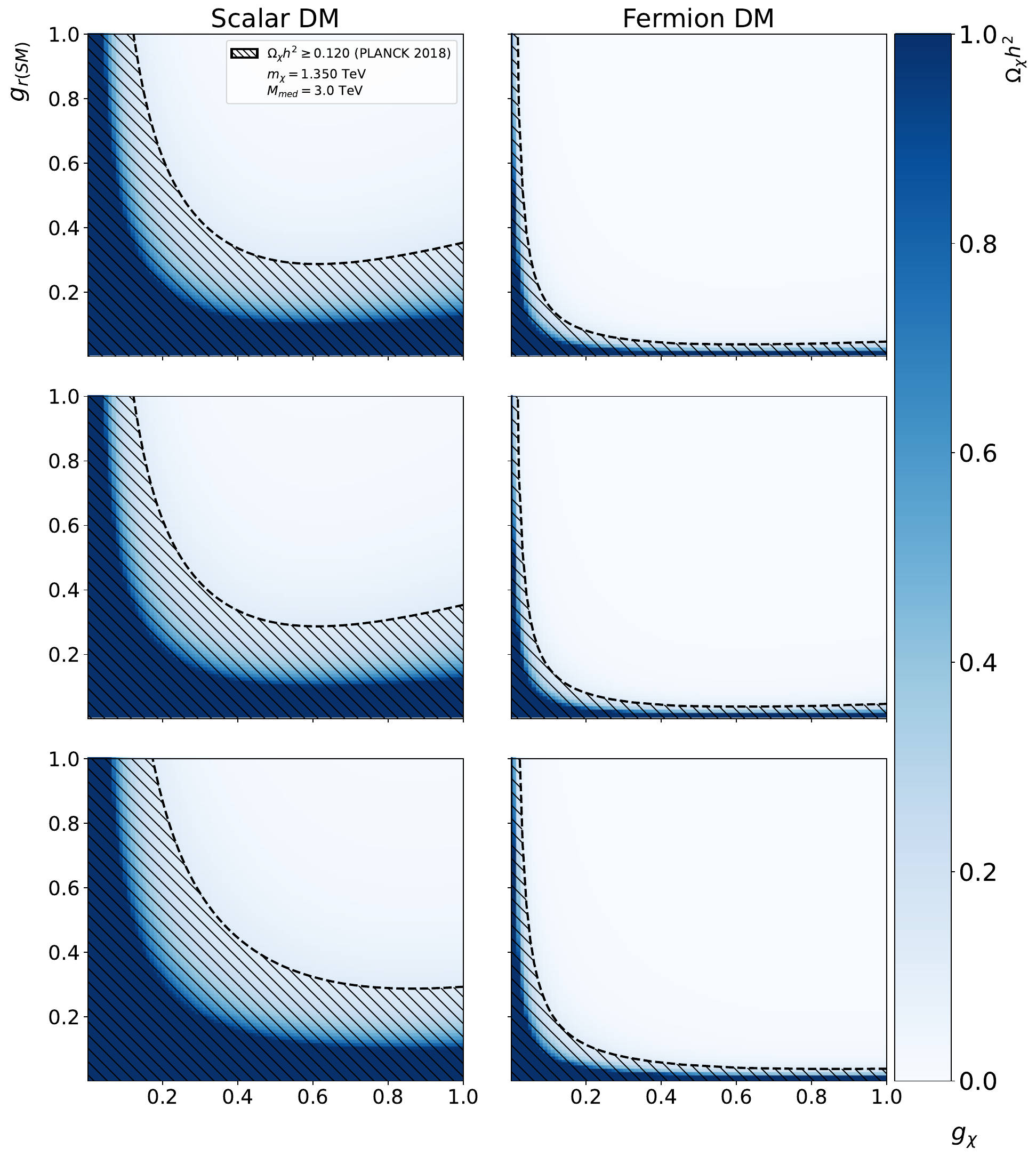}
\caption{\label{omega-relic_coup_qq}Same as \autoref{omega-relic_coup_ee}, but for $pp$ collisions at 14~TeV.}
\end{figure}

Figures~\ref{omega-relic_coup_ee} and \ref{omega-relic_coup_qq} present the scans for couplings $\gr\times\gx$, showing very similar exclusion regions for either the DM species and coupling possibilities as stated in \autoref{couplings}. One can see that the excluded regions in $\ee$ and $pp$ collisions are quite different. The region of overabundance for $\ee$ collisions seen in \autoref{omega-relic-fig_ee} occurs because of the smaller value of $\gr=0.01$ chosen for the $\zp$ coupling to SM leptons, which is largely accessible with varying the DM and mediator masses. The same does not happen for the LHC energy regime, where the $\gr=0.1$ lies close to the limit of overabundance of the relic density.

\begin{table}[!b]
\centering
\caption{\label{couplings} Categorization of SM couplings following the definition given in Ref.~\cite{Schmeier:2013kda}.}
\begin{tabular}{llc}
\textbf{Coupling type} &  & \textbf{Definition} \\ \hline
Vector      &  & $\gl=\gr$  \\
Axial-Vetor &  & $\gl=-\gr$ \\
Right ($\!$\textit{chiral}) &  & $\gl = 0$ 
\end{tabular}
\end{table}

\section{Results and Discussion}
\label{results}

The DM production is very distinctive between lepton and hadron colliders given the available beam energies and detector coverage. We investigate the feasibility of DM production at CLIC and the LHC considering the calorimeter detectors planned/available for photon isolation and reconstruction. Hence, we scan the parameter space and find exclusion regions based on the relic density abundance, which shows possible scenarios in the dark sector by means of the sensibility for different species of DM particles. One of the advantage of using particle colliders for the DM searches is that detectors may be designed to be \textit{multipurpose}, that is, they make it possible to measure a large number of observables within the expected production processes. Furthermore, the high integrated luminosity ($\int{\mathcal{L}}\dif t$) achieved in such colliders reduce the statistical uncertainties for the search of evidence of New Physics. This huge number of events comes together, however, with a large number of background events, but may be subtracted from experimental data with a set of selection criteria and good control of uncertainties and systematic errors, which can be simulated and studied separately \cite{Bauer:2017qwy}.

\subsection{Kinematics in lepton and hadron colliders}
\label{kineepp}

\begin{figure}[!t]
\centering
\includegraphics[width=0.5\textwidth]{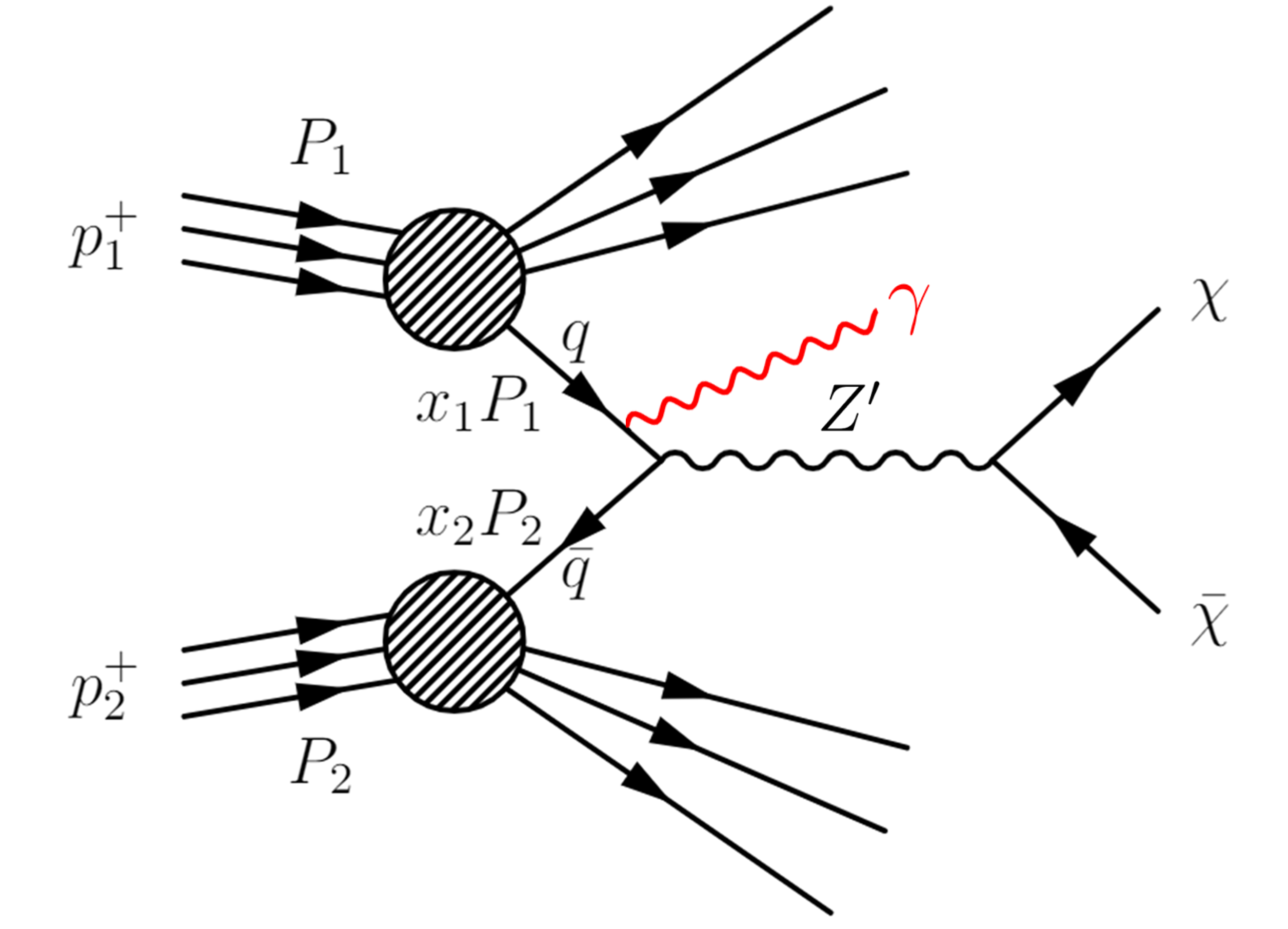}
\caption{\label{drell-yan-DM}A representation of a Drell-Yan-like process mediated by a massive resonance and decaying into DM final states. The same representation is used for different DM species; here shown for the fermion DM case.}
\end{figure}

We start investigating the partonic cross section of DM particle production (\autoref{totalXsec}) in $\ee$ annihilation at CLIC \cite{Dannheim:2012rn,Brunner:2022usy} at $\sqrt{s}=3$~TeV and next $pp$ collisions at the LHC at $\sqrt{s}=14$~TeV, with the initial state fermion mass $m_\psi=m_e$ and $m_\psi=m_{\rm up}$, respectively. The predictions for CLIC are straightforward given the beam-beam annihilation and resonance production. The mono-photon production in $\ee$ collisions is easily obtained by the convolution of the partonic cross section and the photon in ISR,
\begin{eqnarray}
\xstotl(\ee\to\zp\gamma\to\gamma\chi\bar\chi) = \hxstot(\psi\bar\psi\to\zp\to\chi\bar\chi)(1+\delta).
\end{eqnarray}
On the other hand, one needs to employ collinear factorization for a typical Drell-Yan-like process to evaluate the cross section in $pp$ collisions, as shown in \autoref{drell-yan-DM}. In this framework we have for the cross section given by
\begin{equation}
\xstot = \int\limits_0^1 \int\limits_{\tau/x_1}^1 P_{q,\bar{q}}(x_1,x_2)\hxstot(q\bar{q}\to\zp\to \chi\bar\chi)\delta(\tau s-\Mdf^2)\,\dif x_2\dif x_1,
\end{equation}
where $x_i$ are the longitudinal momentum fractions of the proton carried by the partons with $\tau=x_1x_2=\Mdf^2/s$ and
\begin{equation}
P_{q,\bar{q}}(x_1,x_2) = \sum\limits_{q=1}^{N_f} \left[ {f_q({x_1,Q^2}) f_{\bar{q}}(x_2,Q^2) + f_{\bar{q}}({x_1,Q^2}) f_q(x_2,Q^2)}\right],
\end{equation}
is the probability that each quark has a fraction $x_i$ of the total proton momentum. In this work we use the parametrization \texttt{NNPDF31\_lo\_as\_0118} \cite{NNPDF:2017mvq} to model the parton density functions (PDF) within \texttt{LHAPDF} \cite{Buckley:2014ana}, where we consider contributions from $u$, $d$, and $s$ quarks. Considering that we are interested in the cross section near the resonance, we can write the hadronic cross section as function of $\tau$, such as
\begin{equation}
\left. M^2\frac{d\xs}{\dif M^2} \right|_{M=\Mdf} = \tau\int_{\tau}^{1} f_{q}(x_1,Q^2) f_{\bar{q}}(\tau/x1,Q^2)\hxstot(\tau s)\frac{\dif x_1}{x_1},
\end{equation}
with the mass fixed at the dark mediator mass.

The production cross section in $pp$ collisions needs to incorporate the photon ISR within the partonic cross section $\hxstot$ given that the momentum loss by the quark after the photon emission has to be taken into account. Thus, the corresponding production cross section in $pp$ collisions has the form:
\begin{eqnarray}\nonumber
\xstotl(pp\to\zp\gamma\to\gamma\chi\bar\chi) &\equiv& \left. M^2\frac{d\xs}{\dif M^2} \right|_{M=\Mdf} = \tau\int_{\tau}^{1} f_{q}(x_1,Q^2) f_{\bar{q}}(\tau/x1,Q^2)\hxstotl(\tau s)\frac{\dif x_1}{x_1},\\ \\
\hxstotl(q\bar{q}\to\zp\gamma\to\gamma\chi\bar\chi) &=& \hxstot(q\bar{q}\to\zp\to\chi\bar\chi)(1+\delta).
\end{eqnarray}

\begin{figure}[!t]
\includegraphics[width=0.5\textwidth]{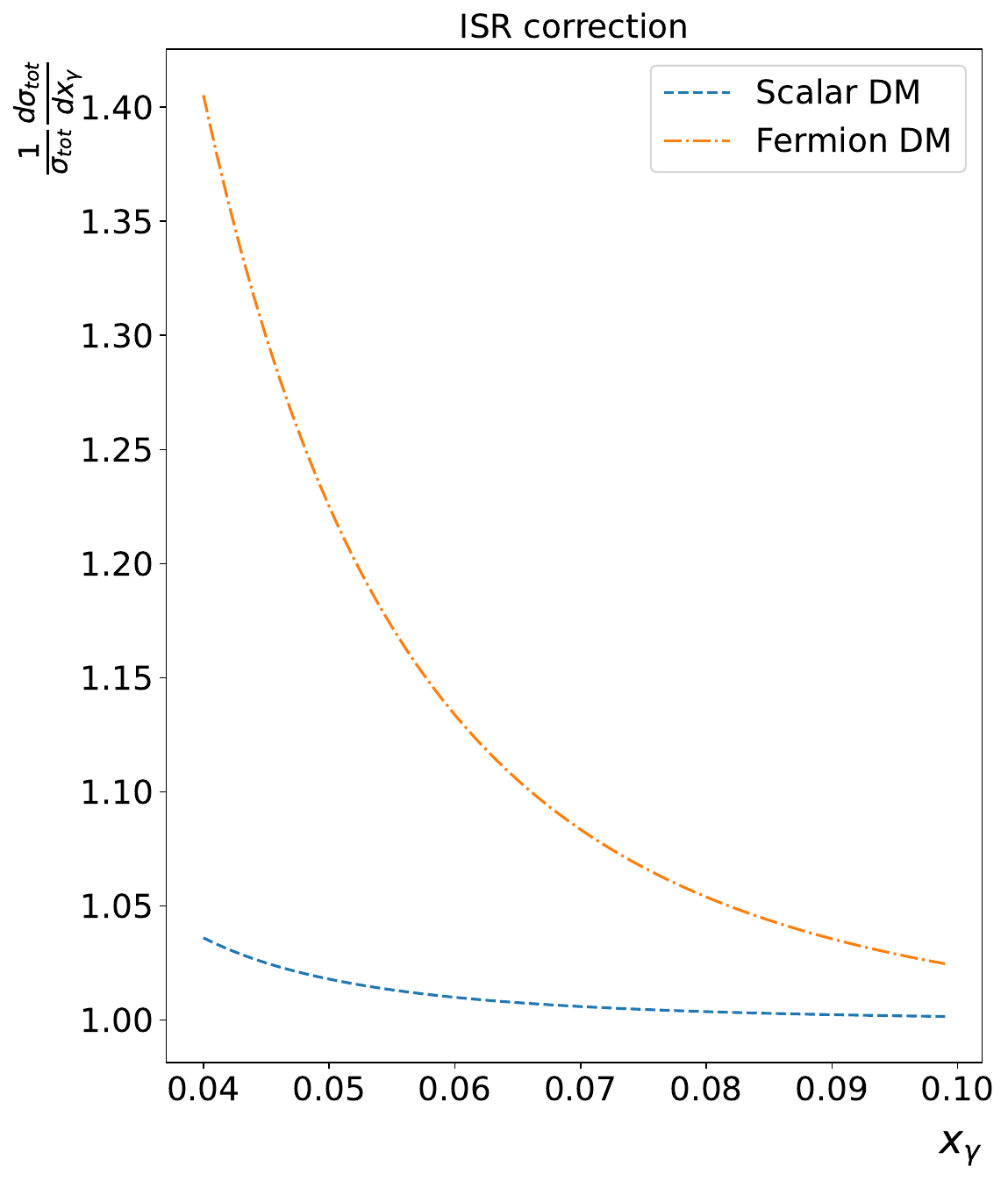}
\caption{\label{dqdsigma}Differential distributions for the ISR factorized cross section as function of the photon energy fraction $\xg=q_\gamma/E_{beam}$, normalized by the total integrated cross section. Here we use $\Mdf=3$~TeV and $\mx=1350$~GeV.}
\end{figure}

We are interested in the SM signal coming from the mono-photon production mechanism in order to observe such an event in the electromagnetic calorimeters. Thus, the photon spectrum is the main experimental signature for searching the DM production. In \autoref{dqdsigma} we show the normalized differential cross section for both $\ee$ and $pp$ collisions as function of the photon energy fraction for the three DM species considered in this work. One can clearly see that the fermion DM case produces a harder photon spectrum than the scalar DM case within the photon energy range near the resonance, which could be a hint for DM production as a experimental signature. Beyond 0.1 the normalized cross section tends to 1.

\subsection{DM parameter scan}
\label{dmscan}

\begin{figure}[!t]
\includegraphics[width=1\textwidth]{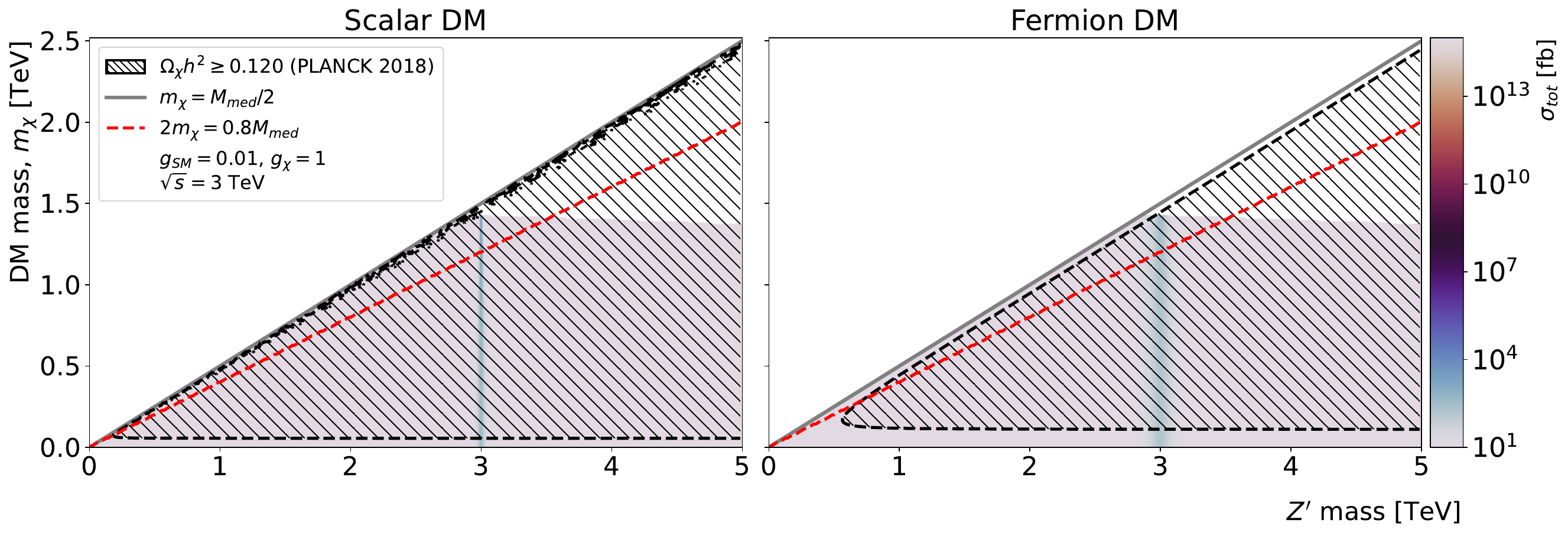}
\caption{\label{massxmassee}Scan of the total cross section with ISR contributions as a function of the particle masses ${\chi}$ of DM ($\mx$) and boson $\zp$ ($\Mdf$) for processes involving $\ee$ collisions. The red dashed line draws the threshold for probing the mass range near the resonance where the relic density is lesser suppressed.}
\end{figure}

\begin{figure}[!t]
\includegraphics[width=1\textwidth]{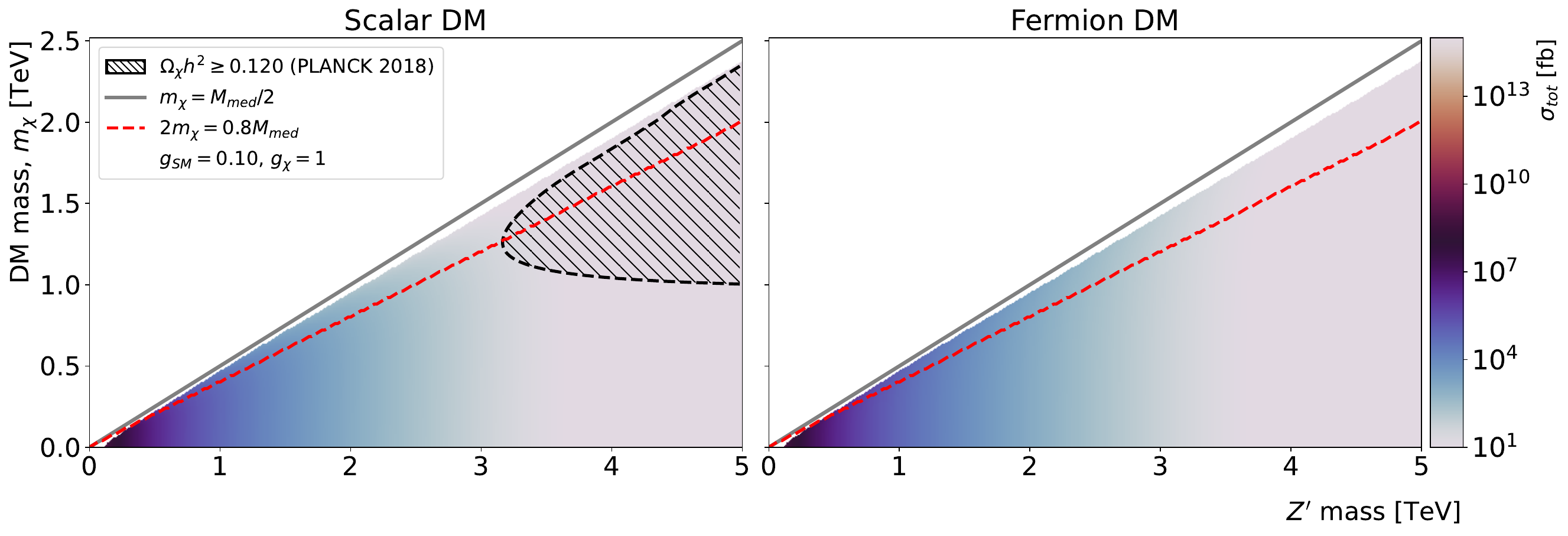}
\caption{\label{massxmassqq} Same as \autoref{massxmassee}, but for $pp$ collisions at 14~TeV and $\gr=0.1$.}
\end{figure}

As we are mainly interested in the DM observation via production by resonances in the $s$-channel, Figures~\ref{massxmassee} and \ref{massxmassqq} show the mass scan in terms of the DM production near the mediator resonance of mass $\Mdf$ decaying into DM particles of mass $\mx$ for both $\ee$ and $pp$ collisions. Given that CLIC will operate with a fixed energy, we scan the masses in Figure~\ref{massxmassee} with the resonant cross section with $\sqrt{s}$ at the $\zp$ peak at 3~TeV and varying $\mx$. Apart from the kinematical limit of $\mx\leq\Mdf/2$, the photon ISR limits the grow of the cross section with $\mx$, producing a slightly decreasing upper bound in $\mx$ beyond $\Mdf > $~3~TeV. Instead, LHC probes different invariant masses and the cross section in Figure~\ref{massxmassqq} is then free to vary with both $\Mdf$ and $\mx$, and the restriction imposed but the ISR appears as a decreasing cross section right below the diagonal for $\mx=\Mdf/2$. As stated before, the coupling of the mediator with the SM is fixed with $\grl = 0.01(0.1)$ for the $\ell\ell\zp$($q\bar{q}\zp$) coupling and $\gx=1$.

The hatched areas show the regions with relic overabundance and thus excluded. The red dashed line show the region with a minimum $\mx$ in agreement with the relic calculation near the $\zp$ resonance, where we note that the cross section for scalar and fermion DM are of the same order of magnitude. As we can see in \autoref{massxmassee}, almost all the mass region is excluded for the CLIC energy regime by the relic overabundance, leaving a tiny mass range available very close to the $\zp$ peak. The results for the LHC, on the other hand, show a much lesser stringent exclusion region or both scalar and fermion DM. Hence, the allowed region above the red dashed line results in a fully available $\mx$ for fermion DM, but a restricted mass range for scalar DM, mostly below $\Mdf$ of 4~TeV.

Considering the potential for observation at particle colliders, if any, one can see very distinctive possibilities within the DM species. The scalar and fermion DM have a significant cross section for smaller $\mx$ and $\Mdf$ masses. However, according to Refs.~\cite{CMS:2021far,ATLAS:2021kxv}, massive mediators in the region below 2~TeV are already excluded with a 95\% confidence level, which favors searches for regions of even higher masses. Besides, a reasonable prediction for the number of expected events needs to take into account the efficiency of identifying invisible end states and the impact of background signals, which are neglected in this work given the specifics of each detector. Hence, the observed event rate will be reduced from these predictions, but still competitive for observation.

The predicted cross section for scalar and fermion DM are similar in both $\ee$ and $pp$ collisions, around to 10$^1$--10$^3$~fb, which would be reduced considering detector efficiencies and background rejection to the level applied in current data analyses, however the expected event rate would be still consistent with the lack of observation as reported by the LHC experiments. Furthermore, in the spectrum chosen for the hard photon emitted as ISR, we noticed little variation in the absolute values of the cross section, despite the fact that there is an evident kinematic constraint near the limit $\Mdf=2\mx$, which indicates that, even in the case of a higher order process, this ends up not disfavoring possible future observations.

In both $\ee$ and $pp$ collisions the regions not excluded by the relic density are very distinct and at very much different scales. Regions of DM production consistent with cosmological observations are strongly excluded for scalar DM due to the very nature of the resonant production process applied here according to Ref.~\cite{Gondolo:1990dk}. This can be seen in the $\ee$ collisions at CLIC, while the restriction is much less stringent in $pp$ collisions. Besides, regions of low DM mass are not accessible in $\ee$ collisions -- neglecting the low $\mx$ masses far from the resonance region --, the $pp$ ones can access the higher DM mass region at TeV scale. The cross section, and correspondingly the event rate, is comparable within the scalar and fermion DM, excluding a tiny region at higher DM mass in the scalar DM case, however this exclusion region depends on the range of photon energy which drives how DM mass in reachable in particle colliders.

\begin{figure}[!t]
\includegraphics[width=1\textwidth]{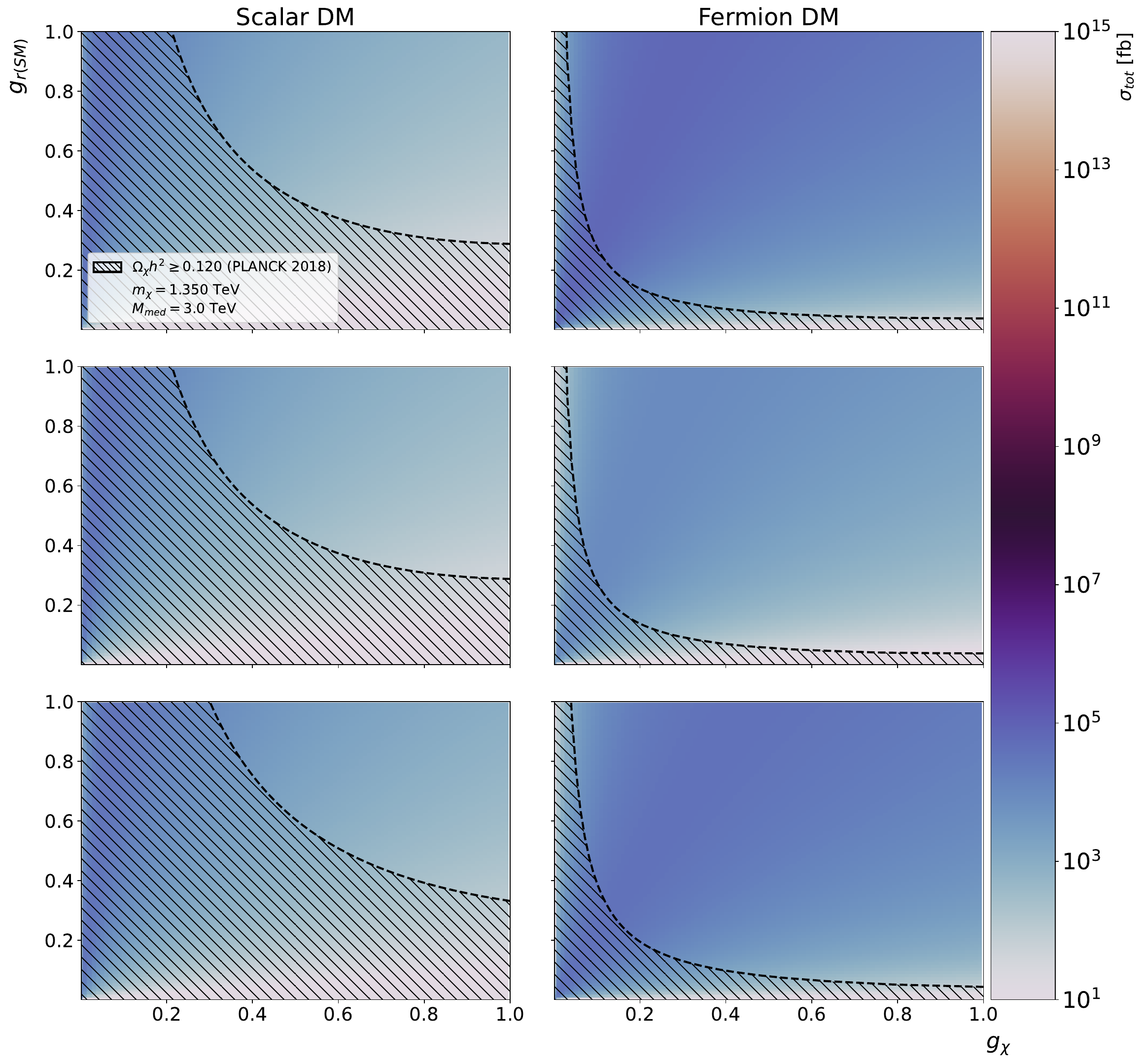}
\caption{\label{coupling-isree}Coupling scan showing the total cross section in $\ee$ collisions at $\sqrt{s}=$~3~TeV for each type of coupling with the SM: vector (top), axial-vector (center), and chiral (bottom). On the horizontal axis we can see the $\gx$ coupling dependence of $\zp$ with the dark sector. The DM particle mass and the mediator are fixed at $\mx=$~1350~GeV and $\Mdf=$~3~TeV.}
\end{figure}

\begin{figure}[!t]
\includegraphics[width=1\textwidth]{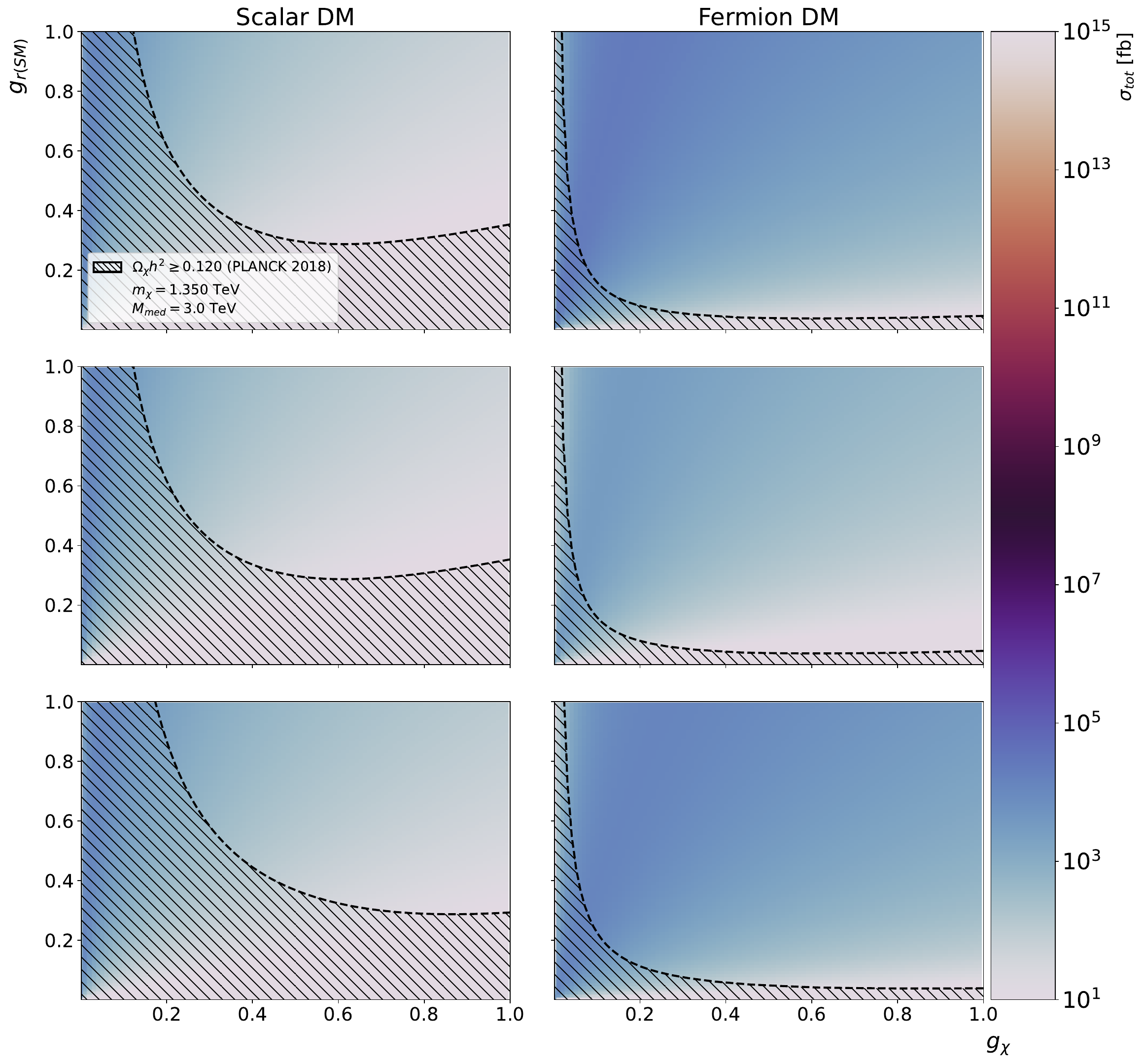}
\caption{\label{coupling-isrpp}Same as \autoref{coupling-isree}, but in $pp$ collisions at $\sqrt{s}=$~14~TeV.}
\end{figure}

Figures~\ref{coupling-isree} and \ref{coupling-isrpp} present the distributions in terms of $\gr$ and $\gx$ couplings with conventional matter and the dark sector for $\ee$ and $pp$ collisions taking into account the photon ISR. We show the cross sections with mediator mass of $\Mdf=3$~TeV and DM mass of $\mx=1350$~GeV so that we are able to see more clearly the regions where the DM mass near the resonance allows the DM relic abundance production to conform with cosmological limits. We can notice that a large part of the phase space for scalar DM is excluded by the relic density, while the fermion DM is the one that presents the best observation opportunity due to its higher cross section in the non-excluded region. In addition, we see a small effect by changing the type of coupling between the SM particles and the dark mediator, potentially because of the small contribution of the coupling constants in the cross sections. One can noticed that the chiral coupling results in a slightly larger exclusion region given the nature of the $\gl$ coupling in the cross sections and decay widths. 

\section{Conclusion}
\label{conclusions}

We present a simplified SM extension with a new renormalizable symmetry group $U_{\chi}(1)$ acting as a vector portal for DM, which can be distinguished by the composition of its fields, i.e., scalar or fermion, by two respective possible interaction Lagrangians. We evaluated the relic density near the resonance of the $\zp$ mediator for the first time in the literature, which drives the exclusion regions in the mass and parameter scans. As a result, we are able to define a region where the relic density is lesser suppressed, providing proper exclusion limits for DM production mediated by a $\zp$ vector boson at TeV scale, showing that our simplified model are within the parameter space probed in the experimental and cosmological limits, excluding a significant region of coupling constants.

The results show that the potential for DM observation in $\ee$ collisions is very challenging, with a tiny region still available for fermion DM very close to the resonance. For $pp$ collisions we can see that there is still a promising region for detection of resonances that can serve as a portal for the DM production, especially for fermion DM as a candidate in the LHC energy regime under these assumptions. The calculation of the relic density near a resonance region is an important factor for obtaining predictions for the LHC and we showed that there still available regions for this resonance production in agreement with cosmological and collider constraints. This work aims to further narrow the parameter space, especially the mass range of the mediator mass, to establish grounds for the search of massive mediators in the resonant $s$-channel production. The search for New Physics, specifically the production of DM in $pp$ colliders, is promising, even in regions already covered by the energy and luminosity of the LHC as well as in future accelerators.

Differently from what happens in direct and indirect searches, if any DM trace have been detected in collider experiments, it would not be possible to state that the observed DM would be the same that has its gravitational effects observed at cosmological levels. This is because the time of flight of a particle to traverse all detector dependencies is not comparable to the cosmological lifetime of a stable primordial DM particle. In addition, cross-analysis of data from different experiments almost always takes some dependence on specific DM models, due to the difficulty of comparing such results independently \cite{Trevisani:2018psx,ParticleDataGroup:2022pth}. Furthermore, the experimental viability of DM observation through these processes are focused in the use of disappearing tracks, as already done by the experiments at theLHC~\cite{CMS:2020atg}. Therefore, our results show that searches of mono-photon production with large missing energy can be a competitive experimental signature for observing evidence of DM particles at the LHC, with the potential of characterizing the nature of a DM vector mediator in production near its resonance. 

\begin{acknowledgments}
This work was partially financed by the Brazilian funding agencies FAPERGS, CAPES and CNPq. This study was financed in part by the Coordena\c{c}\~ao de Aperfei\c{c}oamento de Pessoal de N\'{\i}vel Superior - Brasil (CAPES) - Finance Code 001. GGS
acknowledges funding from the Brazilian agency Conselho Nacional de Desenvolvimento Científico e Tecnológico (CNPq) with grant CNPq/311851/2020-7.
\end{acknowledgments}

\bibliographystyle{apsrev4}
\bibliography{refs}

\end{document}